# MAGNETO-CONVECTION AND LITHIUM AGE ESTIMATES OF THE $\beta$ PICTORIS MOVING GROUP

J. MACDONALD AND D. J. MULLAN
Department of Physics and Astronomy, University of Delaware, Newark, DE 19716
jimmacd@udel.edu, mullan@udel.edu

ABSTRACT

Although the means of the ages of stars in young groups determined from Li depletion often agree with mean ages determined from Hertzsprung – Russell diagram isochrones, there are often statistically significant differences in the ages of individual stars determined by the two methods. We find that inclusion of the effects of inhibition of convection due to the presence of magnetic fields leads to consistent ages for the individual stars. We illustrate how age consistency arises by applying our results to the $\beta$ Pictoris moving group. We find that, although magnetic inhibition of convection leads to increased ages from the Hertzsprung – Russell diagram isochrones for all stars, Li ages are decreased for fully convective M stars and increased for stars with radiative cores. Our consistent age determination for the $\beta$ Pictoris moving group of 40 Myr is larger than previous determinations by a factor of about two.

*Online-only material:* color figures

## 1. INTRODUCTION

In principle, stellar age and mass determinations by different methods should be consistent. However, there are a number of young stars for which results obtained from their position in the Hertzsprung – Russell (H–R) diagram are in discordant with their Li abundances.

Many studies of young open clusters find that the Li age is greater than the age calculated from fitting the upper main-sequence (e.g. Barrado y Navascués, Stauffer, & Patten 1999; Stauffer et al. 1999; Jeffries & Oliveira 2005; Manzi et al. 2008). These results are used to argue that overshooting of the formal boundary of the convective core is occurring in upper main sequence stars, since larger cores increase main sequence life times and consequently the ages derived from upper main-sequence fitting. Convective core overshooting does not change the cluster's Li age which is determined from the degree of Li depletion in the low mass members of the cluster.

A few studies of individual stars have found that their Li ages are older than ages derived from the H–R diagram. For example, Song et al. (2002) found that the lithium depletion age of the M4/4.5 binary HIP 112312 is >20 Myr. They also found that this system is likely a member of the $\beta$ Pictoris Moving Group (BPMG), which has an age of ~12 Myr based on pre-main sequence H–R diagram isochrones (Zuckerman et al. 2001). Similarly White & Hillenbrand (2005) found that the binary T Tauri star StHA 34 has an H–R diagram age of 8 ± 3 Myr whereas its lithium depletion implies an age of about 25 Myr.

Mentuch et al (2008) have determined ages of five young groups of stars from lithium depletion. In general, they find ages for the groups that are consistent with those determined from isochrones for pre-main sequence evolutionary tracks, and dynamical expansion. However stars in individual groups can have a large spread in Li age. For example, stars in BPMG have age determinations ranging from 10 to 30 Myr. More recently Yee & Jensen (2010, hereafter YJ) have independently determined lithium abundances for a number of stars in BPMG. They find that evolutionary models predict much less lithium depletion for the BPMG given its nominal H–R diagram age of ~12 Myr (Zuckerman et al. 2001). This is particularly true for the mid-M stars in the sample, which have no observable Li I line. In general, the ages determined from lithium depletion isochrones are systematically older than from the H–R diagram. YJ suggest that this age discrepancy is related to another troubling discrepancy; that between measured radii of young active M-dwarfs and the smaller radii predicted by evolutionary models (see e.g. Torres et al. 2006; Clausen et al. 2009).

A possible resolution to the radius discrepancy is that the magnetic fields present in active stars acts to impede the flow of energy by convection (Mullan & MacDonald 2001). This leads to larger radii, smaller luminosities and lower effective temperatures. Here we consider how magneto-convection affects the rate at which Li is depleted.

In section 2, we briefly review our model of magneto-convection. In section 3, we illustrate how magneto-convection affects the Li and H – R diagram isochrones. In section 4, we consider specific application of our magnetic models to BPMG. Discussion and conclusions are given in section 5.

2. MODEL FOR MAGNETO-CONVECTION

Based on an energy principle, Gough & Tayler (1966) found that convective stability in the presence of a vertical magnetic field $B_v$ frozen in an ideal gas of pressure $P_{gas}$ and infinite electrical conductivity is ensured as long as $\nabla_{rad}$ does not exceed $\nabla_{ad} + \delta$. Here the magnetic inhibition parameter $\delta$ is defined (in Gaussian cgs units) by

$$\delta = \frac{B_v^2}{B_v^2 + 4\pi\gamma P_{gas}} \quad (1)$$

where $\gamma$ is the first adiabatic exponent.

The Gough – Tayler criterion is not directly applicable to convection in a cool magnetic dwarf for two reasons: 1) The gas is far from ideal and 2) the electrical resistivity is much higher than for fully ionized plasma. The finite magnetic diffusivity allows fluid to move across magnetic field lines, and this can weaken the ability of the magnetic field to hinder thermal convection. To take into account non-ideal thermodynamic behavior and the effects of finite electrical conductivity, Mullan & MacDonald (2010) modified the Gough & Tayler criterion for convection to

$$\nabla_{rad} - \nabla_{ad} > \Delta \equiv \frac{\delta}{\theta_e} \min\left(1, \frac{2\pi^2 \gamma \kappa}{\eta \alpha^2}\right). \quad (2)$$

Here the non-ideal behavior is accounted for by the inclusion of a dependence on the thermal expansion coefficient $\theta_e = -\partial \ln \rho / \partial \ln T|_P$. The factor $\min\left(1, \frac{2\pi^2 \gamma \kappa}{\eta \alpha^2}\right)$, in which $\kappa$ is the thermal conductivity, $\eta$ is the magnetic diffusivity, and $\alpha$ is the ratio of mixing length to pressure scale height, gives a smooth transition to a generalization (Cowling 1957) of the criterion derived by Chandrasekhar (1961, and references therein) for the onset of thermal convection in the presence of a vertical magnetic field in a thermally conducting and magnetically diffusive incompressible fluid.

In this work, we have used eq. (2) as the criterion for onset of convective stability. To determine the convective energy flux, we replace $\nabla_{ad}$ by $\nabla_{ad} + \Delta$ everywhere it appears in the mixing length theory.

The magnetic inhibition parameter $\delta$ is a local variable: in general, its numerical value may vary as a function of radial position in a star, and the question arises as to the appropriate choice for the radial profile of $\delta(r)$. For simplicity, we assume $\delta(r)$ = constant. Mullan & MacDonald (2001) have presented arguments in favor of this choice of radial profile.

To apply the magnetic convection criterion, knowledge is needed of the magnetic diffusivity. At low temperatures and/or high density, the dominant contribution to the electrical resistivity comes from electron – neutral collisions, even when neutral species provide a negligible contribution to the pressure. In this case, as discussed by MacDonald & Mullan (2009), an accurate treatment of the number densities of neutral species, particularly atomic and molecular hydrogen, is required. For the purpose of determining the atomic and molecular hydrogen number densities, we have used an extension of the pure hydrogen equation of state of MacDonald (2001) to mixtures of hydrogen and helium. Using the collision cross sections given by Günther et al (1976), the resistivity from electron-neutral collisions in cgs units can be fitted by

$$\frac{1}{\sigma} = \frac{n_H}{n_e} \frac{7.532\, 10^4}{T^{0.059} \left(1 + 1.723\, 10^{-6} T\right)^{2.441}}, \quad (3)$$

where $n_H$ and $n_e$ are the number densities of hydrogen atoms and electrons respectively. We treat each H$_2$ molecule as two H atoms. At low temperatures, the free electrons are mainly provided by the low first ionization potential elements Na, Ca, and K, which are not included in our equation of state ionization balance calculation. We have explicitly added their contribution to $n_e$. For collisions between electrons and charged particles, we use the electrical conductivity calculated by MacDonald (1991) for particles interacting through a static screened Coulomb potential.

## 3. THE EFFECTS OF MAGNETO-CONVECTION ON THE ISOCHRONES

To determine how inclusion of magneto-convection can affect the Li and H – R diagram isochrones, we have constructed evolutionary sequences for stellar masses ranging from 0.03 to 2.0 $M_\odot$ and $\delta$ ranging from 0 to 0.05 in steps of 0.005. Each sequence is begun on the Hayashi track well before the deuterium phase. The H – R diagram isochrones and the $A(Li) - T_{eff}$ isochrones are shown in figures 1 and 2 respectively, for ages 10, 32, and 96 Myr, and two values of the magnetic inhibition parameter. The red and blue lines are for $\delta = 0$ and $\delta = 0.01$ respectively. Here $A(Li) = 12 + \log(N(Li)/N(H))$ and $T_{eff}$ is the effective temperature.

We see that the main effect of magneto-convection is to slow the evolution to the main sequence. The largest effect on the H – R diagram isochrones occurs for $T_{eff}$ between 4000 and 6000 K. For $T_{eff}$ greater than about 6500 K, the stars have relatively small outer convective zones and the magnetic inhibition has a correspondingly smaller effect on the stars interior structure. For $T_{eff}$ less than about 3000 K the isochrones are much less affected by the presence of the magnetic field. This is a result of the lower degree of ionization in the cooler stars allowing the magnetic field to decouple from the matter so that convection is less impeded. Similar effects are seen in the Li isochrones. Because the magnetic field reduces convective heat flow, interior temperatures are lowered and the depletion of lithium proceeds at a slower rate.

The inclusion of magnetic field also has the important effect of pushing the region of maximum Li depletion to lower effective temperatures. This can be seen in figure 3 where age is plotted against $T_{eff}$ for Li abundance $A(Li) = 0$, and three values of the magnetic inhibition parameter. Note that for $T_{eff} < 3800$ K, the Li age estimate can be lower for models with magnetic field than for non-magnetic models. This is possible for fully convective stars, because magnetic inhibition of convection makes stars of a given mass cooler than non-magnetic stars at the same age. Hence to obtain the same effective temperature at a given age requires that the magnetic star be more massive than a non-magnetic star. The more massive stars have higher internal temperatures than their less massive counterparts, and, provided they remain fully convective, will deplete their lithium more quickly.

## 4. APPLICATION TO THE $\beta$ PICTORIS MOVING GROUP

Here we illustrate the trends that result from inclusion of magneto-convection in stellar models of young cool dwarf stars by applying our results to the members of BPMG analyzed by YJ. Adopting the values of $T_{eff}$, $L/L_\odot$, and $A(Li)$ with associated errors given by YJ, we begin by determining ages from the isochrones that pass through the error boxes in the H – R diagram and the $A(Li) - T_{eff}$ diagram assuming that convection is unaffected by magnetic fields. Our age estimates are shown graphically in fig. 4. At this point we have made no attempt to impose consistency of determinations of masses from the two diagrams.

From figure 4, we see that the ranges of the two age estimates overlap for four objects, CD-64 1208, GJ3305, HIP 29964, and HD 358623A. For the other objects, we see that the 4 coolest have Li ages longer than their H – R diagram ages whereas for the remaining 3 objects, the reverse is true. This trend was also noted by YJ. Based on the YJ data, it is not possible to assign a unique age to all the objects in the sample. The H – R diagram ages range from about 10 Myr for the coolest stars to at least 30 Myr for the hottest star HD 155555 A. Its companion HD 155555 B requires an age of at least 40 Myr based on its position in the H – R diagram. The Li ages range from 0 to greater than 50 Myr. Because HD 155555 A shows no signs of Li depletion (relative to our assumed initial value) it cannot be older than 20 Myr. Averaging the ages over the sample, we find a mean H – R diagram age of 18 Myr and a mean Li age of 14 Myr.

Figure 5 shows the same data as figure 4, except that now the models include magnetic inhibition of convection with $\delta = 0.01$. We see that inclusion of magneto-convection brings the H – R diagram and Li age estimates into agreement with each other for all but four members of the BPMG sample. This indicates that inclusion of the additional physical process of magnetic inhibition of convection is a viable mechanism for resolving age discrepancies in young groups and clusters.

In producing the figures 4 and 5, we have considered only the isochrones and have not considered whether the HR diagram and $A$(Li) give consistent estimates for the stellar masses. Enforcing mass consistency reduces the allowed ranges of ages and fewer stars are found to have consistent ages. For example, when the mass constraint is included, consistent age estimates are found only for GJ 3305 when $\delta = 0$. To include the additional constraint of mass consistency, we determine which of our evolutionary tracks pass through the stars' 3 – dimensional error boxes in $T_{eff} - L - A$(Li) space, and record the lowest and highest ages for which each track is inside the error box. For each star we obtain a set of points in Mass – Age space which enclose the region for which self – consistent ages and masses are possible. These points are shown in figure 6.

Figure 7 shows the age range for each star plotted against the central value of its effective temperature range. We see that for most stars an age of about 40 Myr is possible. The biggest deviants are AU Mic for which the lowest age found is 55 Myr, and GJ 3305 for which the highest age found is 32 Myr. In figure 8, we show the allowed range in $\delta$ for each star plotted against effective temperature. The solid circles indicate the mean of the $\delta$ values that permit an age of 40 Myr. The behavior of $\delta$ is suggestive of an increase in field strength at later spectral type.

5. CONCLUSIONS AND DISCUSSION

We have explored how inclusion of magneto-convection in models of young low mass stars changes H – R diagram isochrones and the rate at which $^7$Li is depleted. Because magneto-convection reduces the efficiency of convective energy transport, the stellar models have lower luminosity and interior temperatures than in the absence of magnetic fields. As a consequence stars contract more slowly and $^7$Li is depleted at a lower rate. The degree to which the models are changed by magneto-convection depends on how much of the star is convective and also on the ionization state of the stellar plasma. In the cooler model stars, with $T_{eff}$ less than about 3000 K, the presence of magnetic field has a relatively small effect, because the lower degree of ionization allows the magnetic field to decouple from the matter so that convection is less impeded. The hotter models, with $T_{eff}$ greater than about 6500 K, have relatively small outer convective zones and the magnetic inhibition has a correspondingly smaller effect on the stars interior structure.

We have applied our models to the data obtained by Yee & Jensen (2010) for the $\beta$ Pictoris moving group. By comparing their data with evolutionary model isochrones and Li depletion, Yee & Jensen found an anti-correlation between H – R diagram and Li depletion ages. We find that inclusion of magnetic inhibition of convection leads to age agreement for all stars in their sample, and that a consistent age of 40 Myr is found for all but two stars in their sample. The two discordant stars have age limits that bracket 40 Myr.

Magnetic inhibition of convection has the effect of increasing the radii of low mass stars in thermal equilibrium. At an age of 40 Myr, our magnetic models for the coolest stars in the YJ sample, which we predict to have masses $\approx 0.35\ M_\odot$, have radii that are about 10% larger than non-magnetic models of the same age and mass. For the K stars in the sample the magnetic models are about 7.5 % larger than non-magnetic models. These relative differences are similar to those between observed and theoretical radii of the components of many low mass close binaries. For example, Morales et al. (2009a) find that both components of the M dwarf binary CM Dra are 5% larger than predicted by models. Ribas (2003) finds that the components of the M dwarf binary CU Cnc are 10% larger than models. For K stars, Morales et al. (2009b) find the K7V secondary of IM Vir to be 7.5% larger than predicted by models, and the K3 secondary in UV Psc of mass 0.76 $M_\odot$ and radius 0.83 $R_\odot$ (Popper 1997) is oversized by about 10% compared to our models. This comparable difference in stellar radii provides strong evidence to the idea that discrepancies in ages of clusters and groups from H – R diagram and Li – depletion isochrones is related to components of low mass close binaries being oversized compared to model predictions. We propose that the single resolution to these two problems is inhibition of convection by magnetic fields.

Mentuch et al. (2008) have also derived Li abundances for a number of stars in the $\beta$ Pictoris moving group. Comparing stars common to both samples, the Mentuch et al. Li abundances are in general higher than those of Yee & Jensen by up to an order of magnitude. The major exception is CD -64 1208 for which the Mentuch Li abundance is lower by a factor of 25. However Mentuch et al also find a significantly lower effective temperature than YJ for this star. This difference is possibly related to the rapid rotation of CD -64 1208, $v\sin i = 110$ km s$^{-1}$. We have tested the sensitivity to systematic differences in Li abundance by artificially increasing the YJ values by factors of 3 and 10 (up to a limit of the solar system Li abundance). Since the Li isochrones are almost vertical at low temperatures (see fig. 2), the consistent ages for the cooler stars (with $T_{eff}$ less than about 3700 K) are essentially unchanged. However an increase in Li abundance leads to a decrease in consistent age for the hotter stars. Even so, an age of 40 Myr for most of the moving group remains possible if the Li abundances are 3 times higher than the YJ values.

Our age of 40 Myr is significantly higher than other estimates for the age of the $\beta$ Pictoris moving group, e.g. ~12 Myr (Zuckerman et al. 2001), 21 ± 9 Myr (Mentuch et al 2008). This trend appears at first site to be counter to the finding that Li ages of young clusters, such as the Pleiades, are significantly older than the ages found from fitting the upper main sequence. However for the coolest stars, i.e. those used to determine the lithium ages of young clusters, we find that magnetic inhibition of convection reduces their Li age estimates by factors of 2 or more. We propose this as an alternative solution to the cluster age problem, which removes the need to reduce the age of upper main sequence stars by including core convective overshoot. Interestingly, our method of modeling magneto-convection, which is equivalent to increasing the adiabatic gradient (by the magnetic inhibition parameter $\delta$) is complementary to the method of producing convective overshoot by reducing the adiabatic gradient by some prescribed amount (Schröder, Pols, & Eggleton 1997).

If we adopt an age for BPMG of 40 Myr, our models allow us to predict the masses of the stars. A number of stars in BPMG are members of wide binaries for which mass ratios can in principle be determined astrometrically. One promising system is the K7-M0 binary HD 358623 discovered by Jayawardhana & Brandeker (2001). Our predicted masses are 0.86 ± 0.01 $M_\odot$ for the K7 primary and 0.54 ± 0.03 $M_\odot$ for the M0 secondary. Hence we expect the mass ratio to be 0.63 ± 0.04. Based on the positions in the H – R diagram, our non-magnetic models give an age of 13.5 ± 2.5 Myr for this binary, and component masses of 0.84 ± 0.05 $M_\odot$ and 0.24 ± 0.03 $M_\odot$. The predicted mass ratio is 0.29 ± 0.05. The Li ages from our non-magnetic models, 13 ± 6 Myr for the primary and 31 ± 7 Myr for the secondary, do not agree which illustrates the need to consider magnetic models. The predicted masses are 0.90 ± 0.13 $M_\odot$ and 0.22 ± 0.05 $M_\odot$, so that the mass ratio is 0.26 ± 0.09. In either case the mass ratio is significantly less than for the consistent magnetic model. This binary system, resolvable with adaptive objects, has an orbital period of order 700 years, and hence with a little patience the mass ratio could be obtained. Astrometric determination of the orbit should be easily achieved by an instrument such as GAIA.

We find that to get a consistent age for BMPG the magnetic inhibition parameter needs to be larger for the cooler stars. For $T_{eff}$ > 3700 K, $\delta \approx 0.01$ whereas for $T_{eff}$ < 3400 K, $\delta \approx 0.04$. This division in magnetic properties might be associated with a difference in dynamo mechanism. The stars with $T_{eff}$ > 3700 K all have radiative cores and the magnetic field may be due to a tachoclinic dynamo as is believed to be the case for the Sun (Browning et al. 2006). Stars with $T_{eff}$ < 3300 K are fully convective, even with $\delta \sim 0.04$, and the magnetic field might be due to a distributed dynamo (Browning 2008).

We thank John Gizis for fruitful discussions. The work of JM is supported in part by a grant from the Mount Cuba Astronomical Foundation. The work of DJM is supported in part by the NASA Delaware Space Grant program.

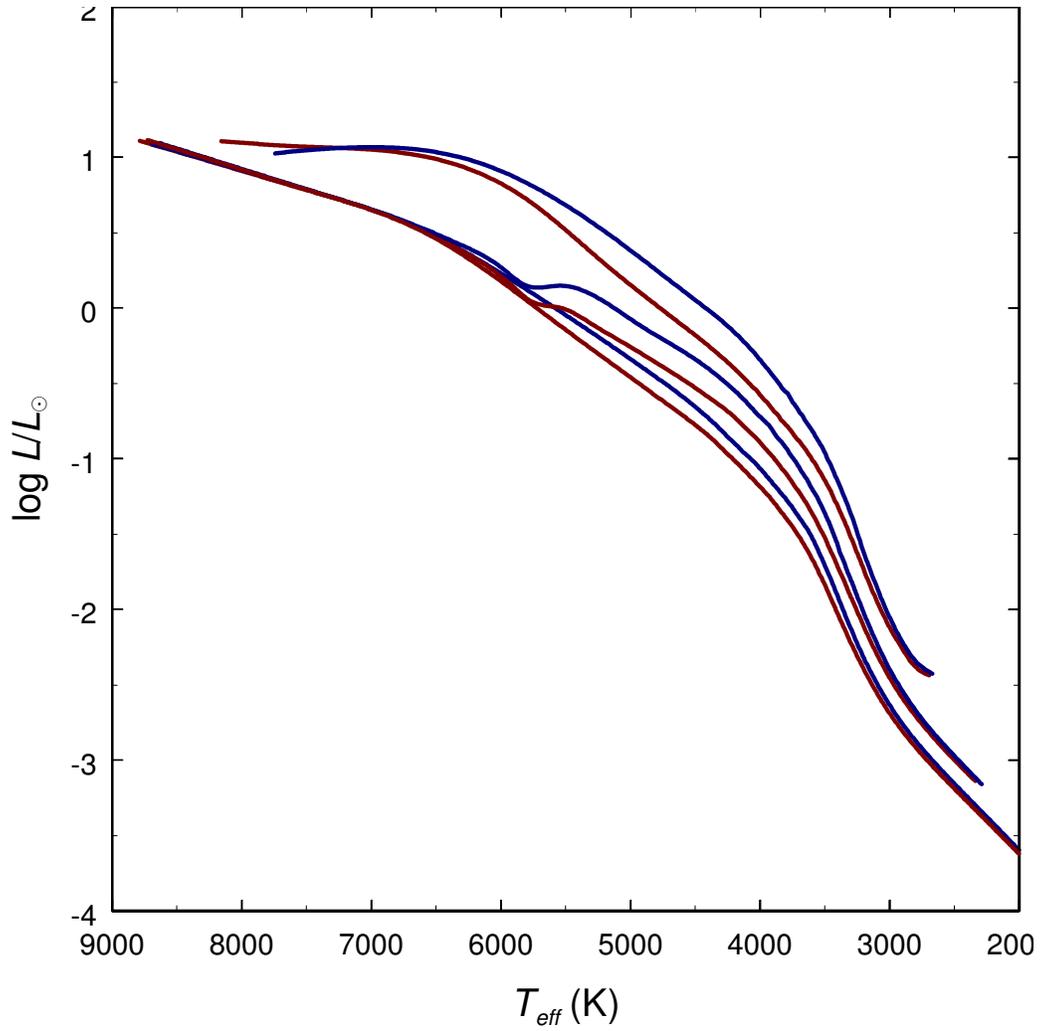

Figure 1. H – R diagram isochrones for magnetic inhibition parameter $\delta = 0$ (in red) and $\delta = 0.01$ (in blue). The ages from top to bottom are 10, 32, and 96 Myr.

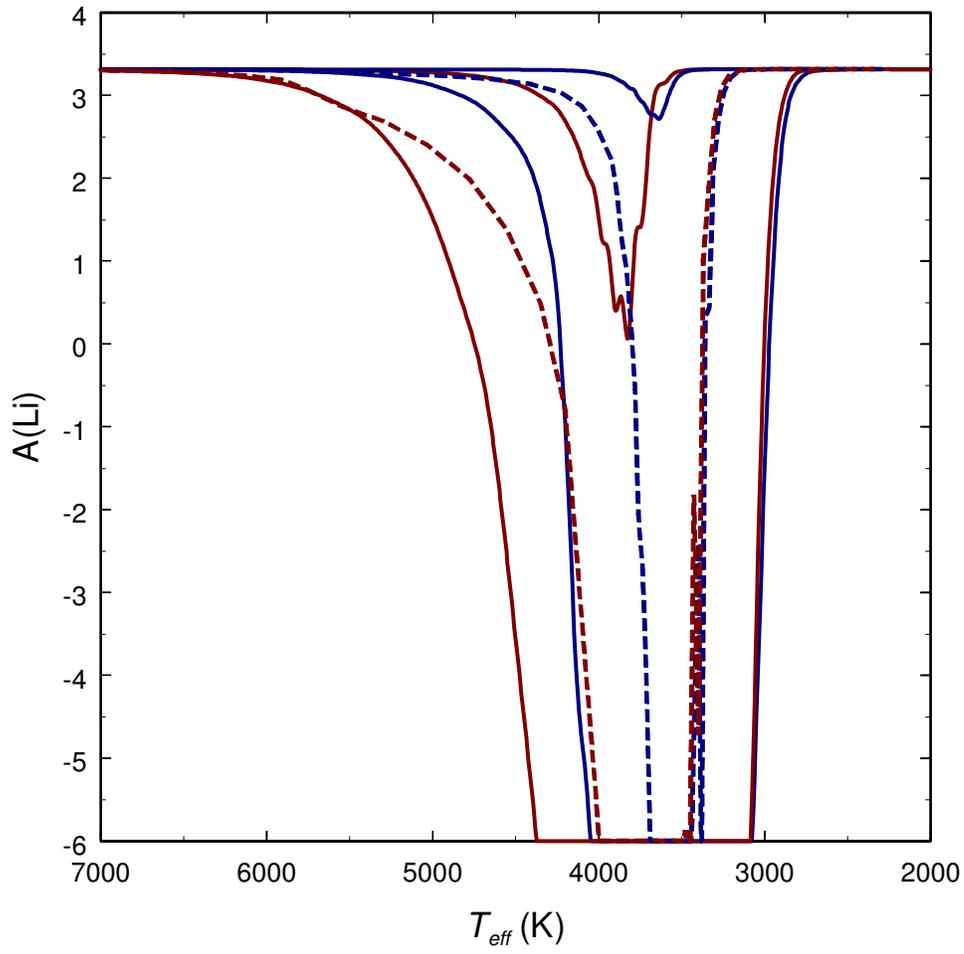

Figure 2. Li isochrones for magnetic inhibition parameter $\delta = 0$ (in red) and $\delta = 0.01$ (in blue). The ages are 10 (upper solid line), 32 (broken line), and 96 Myr (lower solid line).

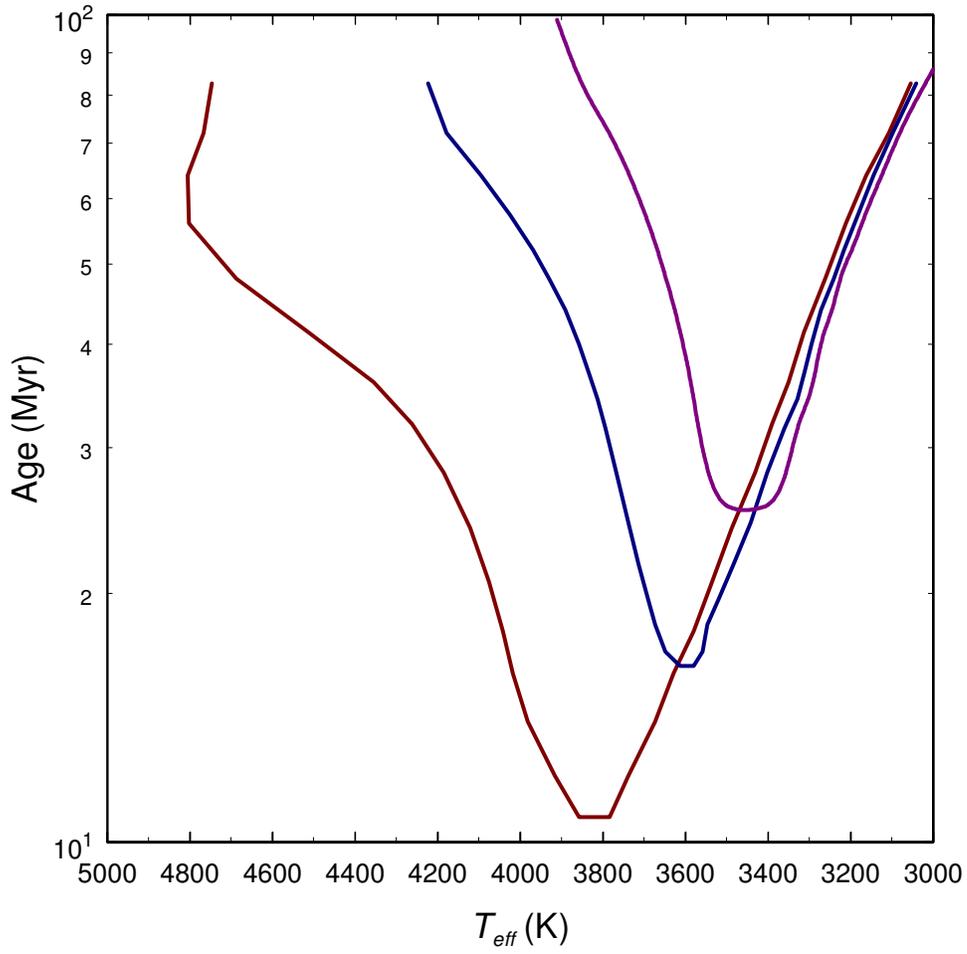

Figure 3. Age at which Li is depleted to $A(Li) = 0$ as a function of effective temperature for magnetic inhibition parameter $\delta = 0$ (red), $\delta = 0.01$ (blue), and $\delta = 0.02$ (magenta).

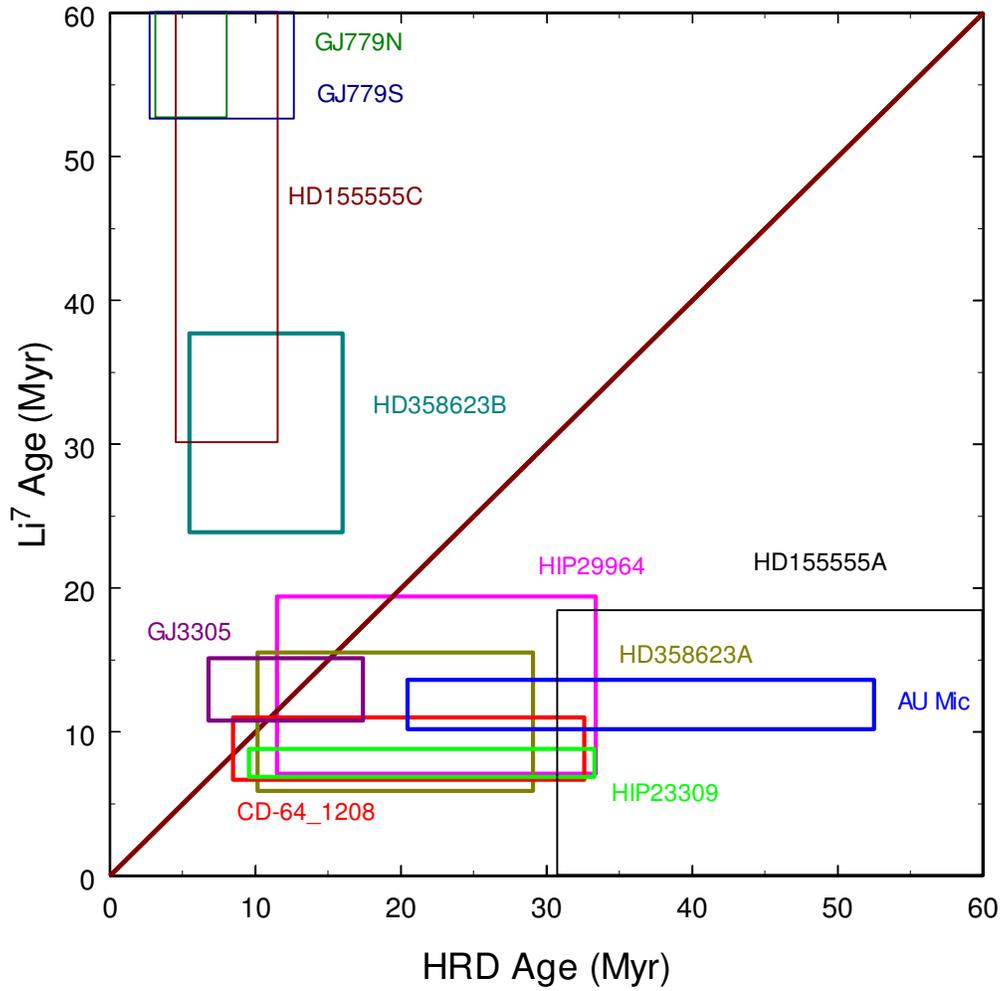

Figure 4. Comparison of age determinations for members of BPMG from the H – R and the $A(\mathrm{Li}) - T_{eff}$ diagrams assuming that convection is unimpeded by magnetic fields.

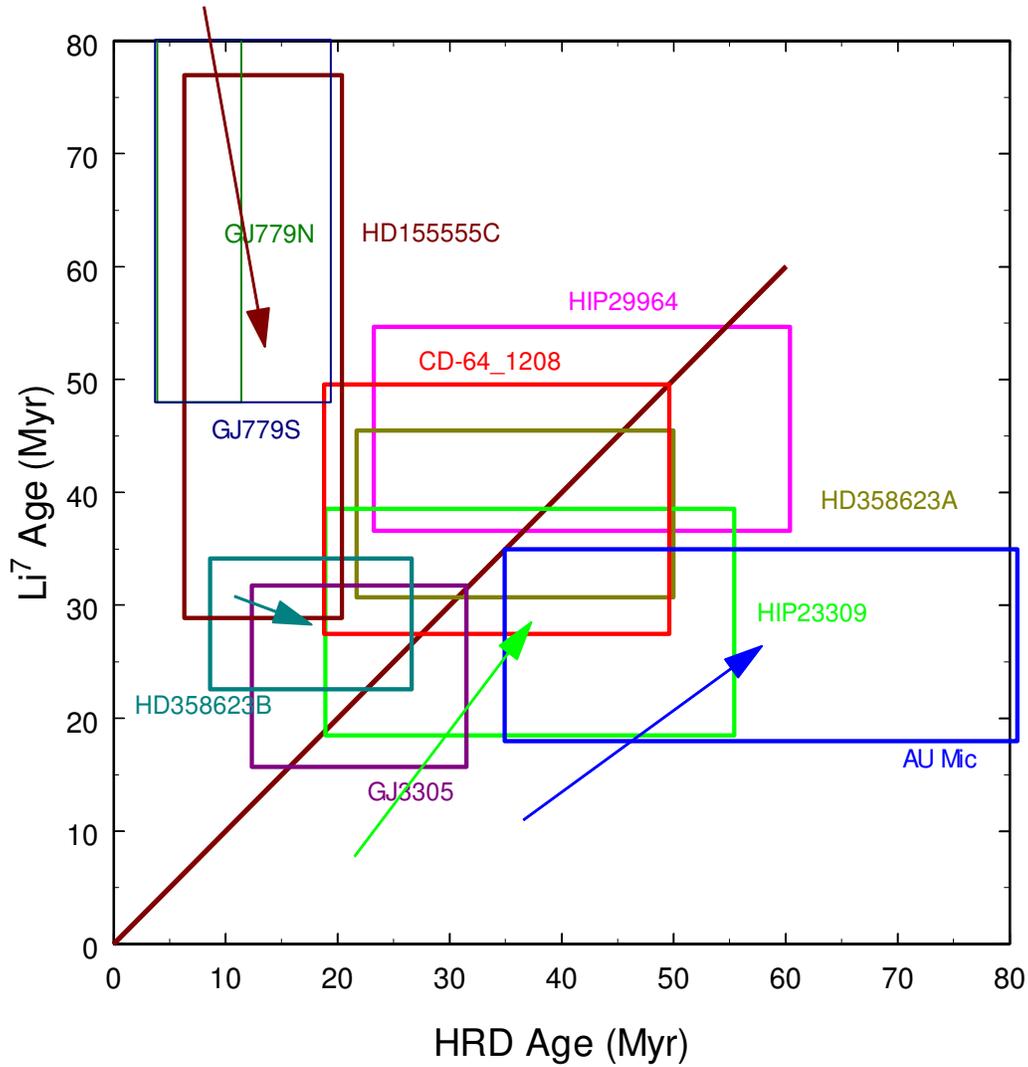

Figure 5. Comparison of age determinations for members of BPMG from the H – R and the $A(Li) - T_{eff}$ diagrams assuming that convection is impeded by magnetic fields corresponding to $\delta$ = 0.01. The arrows show the movement of the center of the error box from its location for $\delta$ = 0.0, for a few representative stars.

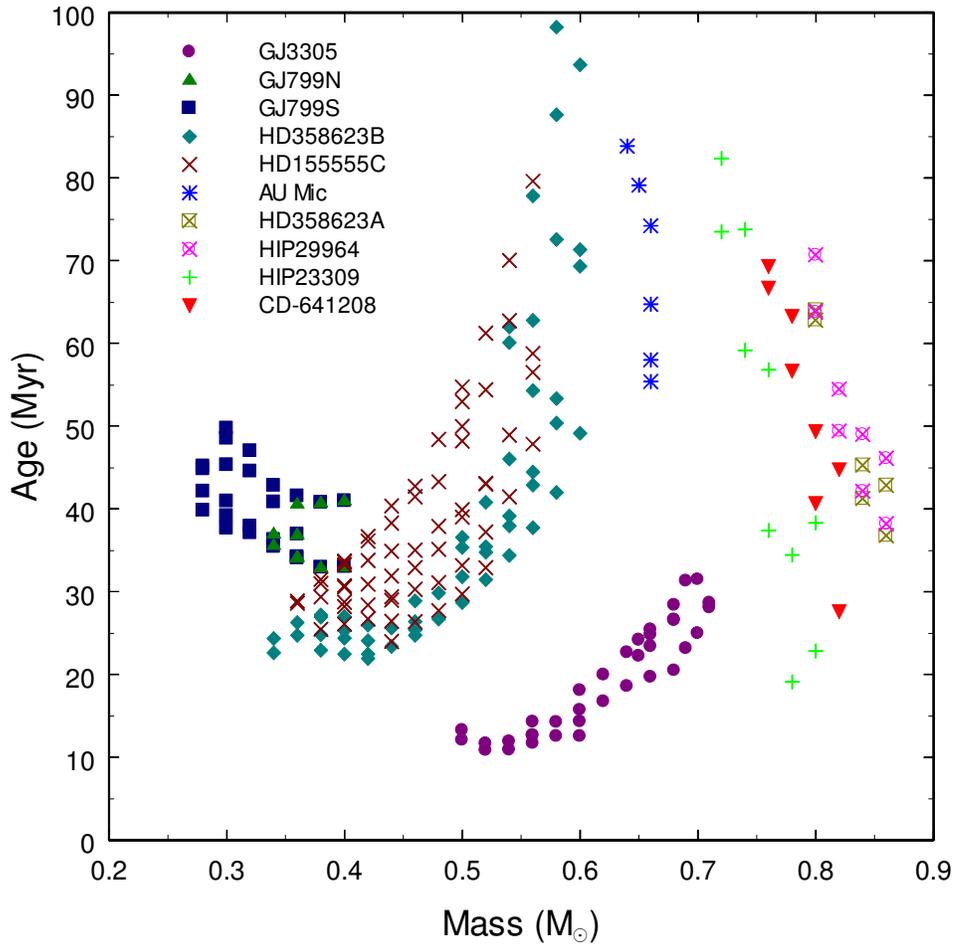

Figure 6. Self consistent age and mass values for BPMG stars in the YJ sample.

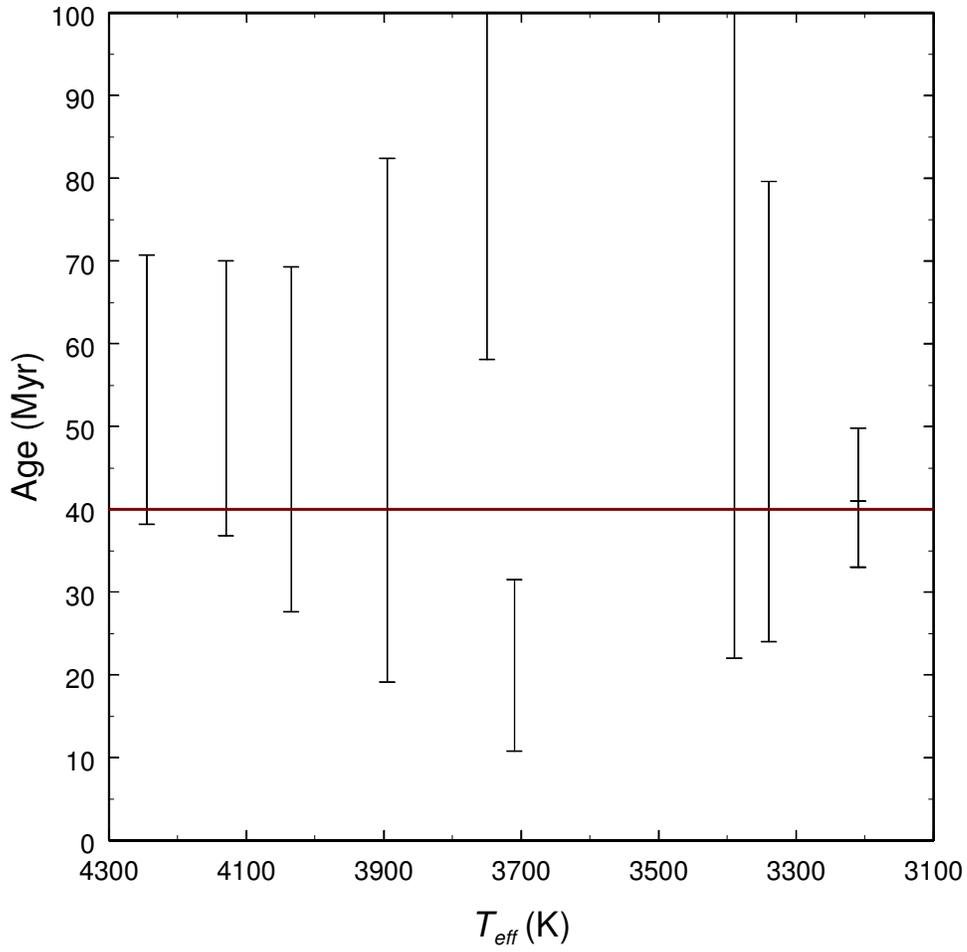

Figure 7. Allowed age range for each star in the YJ sample plotted against its effective temperature. The horizontal line is our favored age of 40 Myr for BPMG

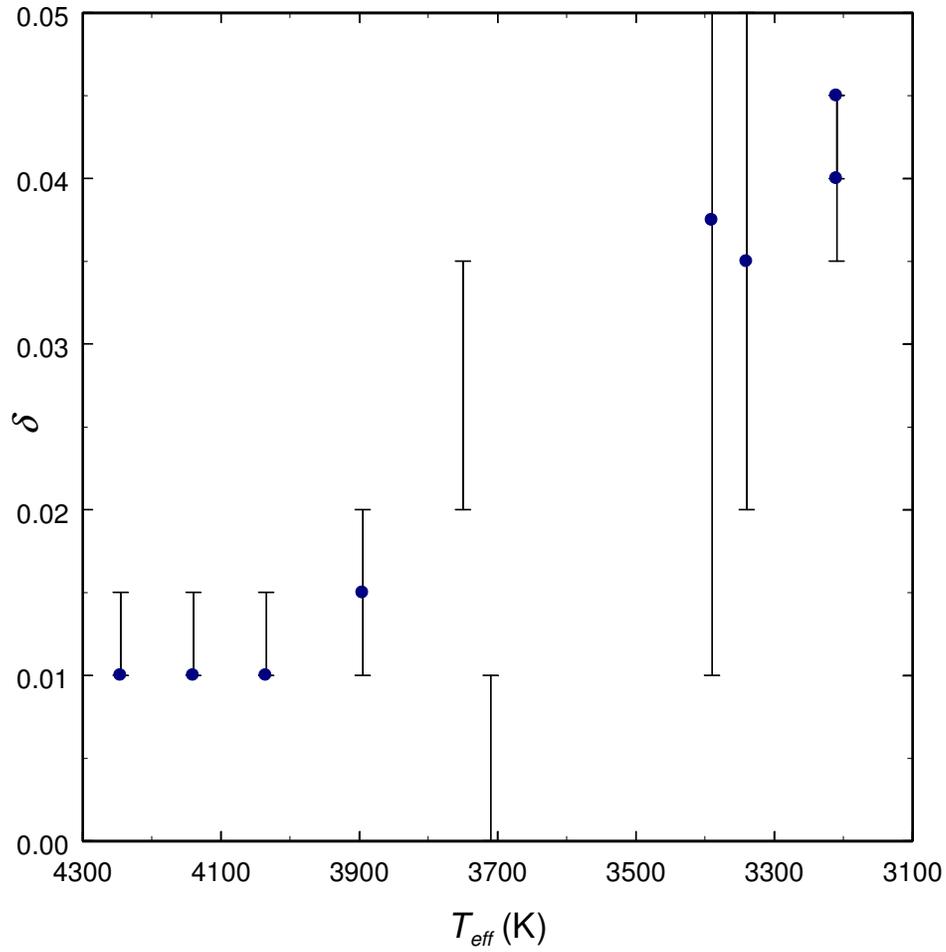

Figure 8. Allowed range in magnetic inhibition parameter $\delta$ for each star in the YJ sample plotted against its effective temperature. The solid circles indicate the mean of the $\delta$ values that permit an age of 40 Myr.